# Eddy currents in accelerator magnets

*G. Moritz*
GSI, Darmstadt, Germany

**Abstract**
This paper covers the main eddy current effects in accelerator magnets — field modification (time delay and field quality) and resistive power losses. In the first part, starting from the Maxwell equations, a basic understanding of the processes is given and explained with examples of simple geometry and time behaviour. Useful formulas are derived for an analytic estimate of the size of the effects. In the second part the effects in real magnets are analysed and described in comparison with numerical and measured results. Finally, based on the previous parts, design recommendations are given regarding how to minimize eddy current effects.

## 1 Introduction

Eddy currents play an important role in magnets. As the title already indicates, this lecture will be restricted to eddy currents in accelerator magnets.

### 1.1 Definition

According to Faraday's law a voltage is induced in a conductor loop, if it is subjected to a time-varying magnetic flux. As a result, current flows in the conductor if there exists a closed current path.

'**Eddy currents**' appear in extended conducting media if these media are subjected to time varying magnetic fields. They are now distributed in the conducting media.

A comprehensive treatment is given in Ref. [1], and Tegopoulos et al.

### 1.2 Effects

We can observe different effects due to these induced eddy currents. They create

- a magnetic field
    - superimposed on the causative time-varying field
    - delayed due to Lenz's law
- electrical power loss
- Lorentz forces, due to the interaction with the magnetic field

#### 1.2.1 Desired effects

Many devices use these effects for their specific applications. Examples are induction heating, magnetic shielding, the levitated train, dampers and brakes [2].

#### 1.2.2 Undesired effects

The above mentioned eddy current effects are mostly undesired in accelerator magnets.

The magnetic field, created by the eddy currents, destroys the high field quality in the 'good field region' that is required for this type of magnet. Moreover the field change is delayed, the field lags behind, and one has to wait for the eddy currents to die out before one can reach a stable situation for the beam. For long time constants (of the order of seconds) this can be a real nuisance and reduce the performance of the accelerator.

Power loss due to eddy currents leads to heating of the (conductive) magnet components, which is especially detrimental for the case of superconducting magnets operating at cryogenic temperatures of a few degrees Kelvin.

Especially in the case of fast changes of high magnetic fields the Lorentz forces must not be neglected. They may lead to stresses in the material beyond the elastic limit.

### 1.3 Goal and outline of the lecture

The primary goal of this lecture is therefore to provide a good understanding of the physics of eddy currents. Analytical formulas will be given to estimate the most common effects under certain assumptions. The results will be compared with numerical calculations.

The effects of eddy currents in different components of a magnet, such as yoke, mechanical structure, coil and beam pipe will be considered with the goal of attaining an optimal magnet design in the presence of a time-varying magnetic field. In most cases this requires a reduction of induced eddy currents as much as possible, in order to minimize their effects.

## 2 Basics

In this section we will develop a basic understanding of the eddy current processes, first by using the diffusion approach and second by analytical solving of Maxwell's equations directly, for some selected examples.

### 2.1 Maxwell equations

In this lecture we consider only the quasi-stationary approach, i.e. neglecting the displacement $d$. That is a good assumption for accelerator magnets with typical frequencies of the order of some hertz. A further assumption is that we have no excess charge ρ.

Then the Maxwell equations look as follows:

Differential form:        Integral form:

Ampere's law: $\quad \nabla \times H = j \quad$ (1) $\qquad \oint H \cdot ds = \int_A j \cdot dA \quad$ (2)

Faraday's law: $\quad \nabla \times E = -\dfrac{\partial B}{\partial t} \quad$ (3) $\qquad \oint E \cdot ds = -\dfrac{\partial}{\partial t} \int_A B \cdot dA \quad$ (4)

$\quad \nabla \cdot E = 0 \quad$ (5) $\qquad \oint_A E \cdot dA = 0 \quad$ (6)

$\quad \nabla \cdot B = 0 \quad$ (7) $\qquad \oint_A B \cdot dA = 0 \quad$ (8)

Material properties: $$B = \mu_0 \underline{\underline{\mu_r}} H \quad (9) \qquad j = \underline{\underline{\sigma}} E \quad (10)$$

The notation is as used in the literature (Knoepfel).

## 2.2 Diffusion approach

In this subsection the eddy current problem is handled as a diffusion problem. The physical background is the well-known Lenz's law: **'the electromagnetic field induced in an electric circuit always acts in such a direction that the current it drives around a closed circuit produces a magnetic field which opposes the change in magnetic flux'.**

The reason for this behaviour is the 'minus' sign in Faraday's law. It is demonstrated in Fig. 1, where a permanent magnet is moved in and out of a solenoidal coil. Please note that the flux created in the coil is opposed to the flux change induced by the movement of the permanent magnet. In other words: the eddy current tries to keep the flux status and consequently delays the change of the external field.

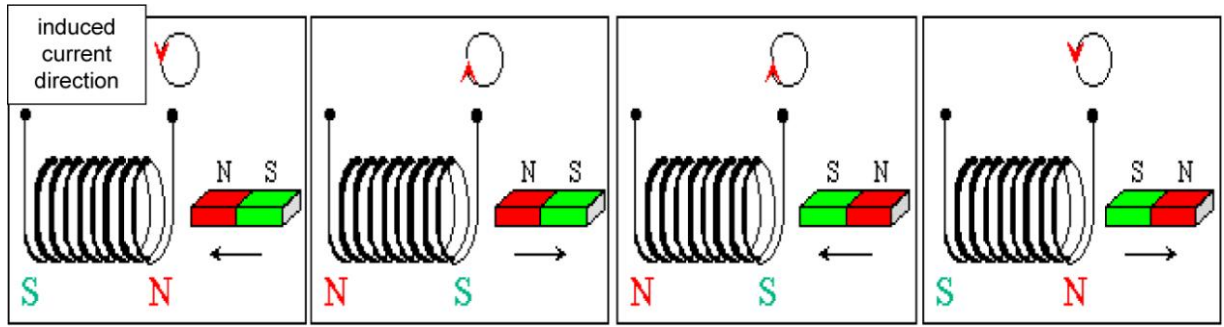

**Fig. 1:** Demonstration of Lenz's law

Therefore we can consider the process as a diffusion process of the magnetic field with a certain time constant, which depends on the electrical properties of the conductive circuit, and in the case of eddy currents on the electrical properties of the extended media. The advantage of the diffusion approach is (besides the clear physical meaning) that the appropriate methods of solution (also the numerical ones) which are well known from the field of thermal diffusion can be used.

### 2.2.1 Field diffusion equation (Knoepfel)

Taking the curl of Ampere's law (1) and introducing the material equation (10) we obtain under the assumption that the conductivity σ is uniform in space:

$$\nabla \times \nabla \times \vec{H} = \nabla \times \vec{j} = \sigma (\nabla \times E) \cdot \quad (11)$$

Application of the well-known vector relation to the left-hand side and introducing Faraday's law (3) on the right-hand side leads to

$$\nabla^2 \vec{H} = \sigma \mu \frac{\partial \vec{H}}{\partial t} \quad (12) \quad \text{with} \quad \mu = \mu_0 \underline{\underline{\mu_r}} \cdot$$

This is the magnetic diffusion equation. The expression κ is called the 'magnetic diffusivity'.

$$\kappa = \frac{1}{\sigma \cdot \mu} \ . \qquad (13)$$

The interpretation is straightforward: the higher the conductivity (i.e., large eddy currents), the lower the diffusivity, i.e., the diffusion process is slow. Large $\mu$ means large stored energy, which takes a longer time to be stored.

Having solved the differential equation for the field $H$ (under certain boundary conditions), the eddy current density can be calculated by Ampere's law and consequently the power loss as well. Diffusion equations can also be derived in a similar way for the magnet vector potential $A$ with the definition

$$B = \nabla \times A \ . \qquad (14)$$

Having found the vector potential by solving the corresponding diffusion equation (under the chosen boundary conditions) one can calculate the current density $j$ by the equation

$$\vec{J} = \sigma \vec{E} = -\sigma \frac{\partial \vec{A}}{\partial t} \ . \qquad (15)$$

### 2.2.2 *Analytical solutions of the diffusion equation (Examples)*

Of course analytical solutions can only be given for simple applications, as far as the geometry of the components, the properties of the materials, and the variation in time are concerned. Nevertheless these solutions give an insight into the physics of the eddy current phenomena and give some simple formula for the estimation of eddy current effects in a magnet.

In the following sections we closely follow the treatment given by (Knoepfel).

#### 2.2.2.1 *Half-space conductor*

The easiest start to understand the diffusion process and the role of the eddy currents in it, is to look at the 'half-space conductor' in one dimension only (Fig. 2).

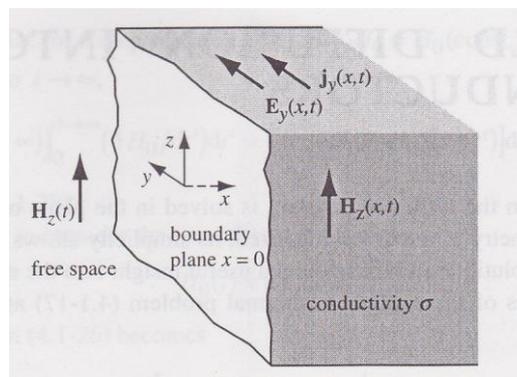

**Fig. 2:** Half-space conductor

In free space an external field $H_z(t)$ is applied. We seek the solution $H_z(x, t)$ of the differential equation

$$\frac{\partial^2 H_z}{\partial x^2} = \sigma\mu \frac{\partial H_z}{\partial t} \quad (16)$$

in the half-space for the specific boundary conditions

$$H_z(0,t) = 0 \quad\quad t < 0$$
$$H_z(0,t) = H_z(t) \quad\quad t \geq 0$$
$$H_z(x, 0) = 0 \quad\quad 0 < x < \infty$$

We shall restrict ourselves to the three most common time dependencies: the step function field, the transient linear field, and the transient sinusoidal field.

2.2.2.1.1  Step function field $H_z(t) = H_0 =$ constant

We define a 'response function' $S(x, t)$ (schematically shown in Fig. 3) by the following relation:

$$H_z(x,t) = H_0 * S(x,t)$$

with $S(x,0) = 0$ and $S(x, t\to\infty) = 1$.

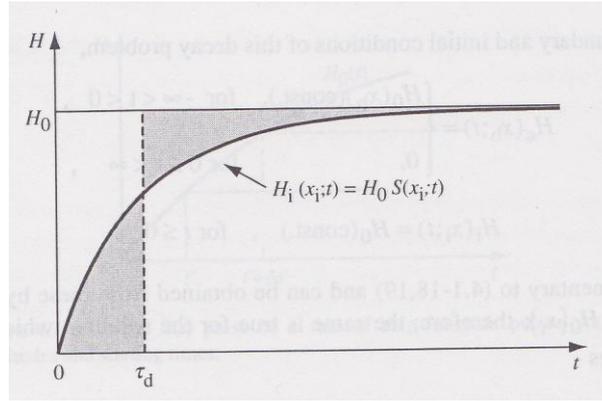

**Fig. 3:** Response function

An average diffusion time constant $\tau_d(x)$ can then be defined by

$$\tau_d = \int_0^\infty \left(1 - S(x,t)\right) dt \,. \quad (17)$$

It looks plausible to introduce a similarity variable $\xi$

$$\xi = \frac{x}{2\sqrt{\kappa t}} \,. \quad (18)$$

With this variable the magnetic diffusion differential equation (16) can be solved. One obtains the following solution

$$H_z(x,t) = H_0 \cdot (1 - \mathrm{erf}\,\xi) \,. \quad (19)$$

with the special response function $S(x, t) = S(\xi) = 1-\mathrm{erf}(\xi) = \mathrm{erfc}\,(\xi)$.

In Fig. 4 is plotted the error function and its complement erfc. Figure 5 shows the dependence of the response function on time, at different coordinates.

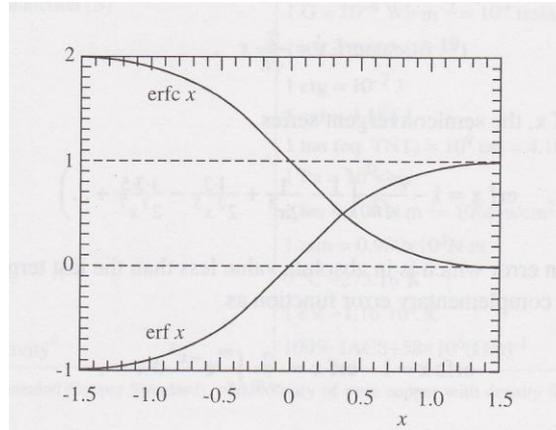

**Fig. 4:** Error function and its complement

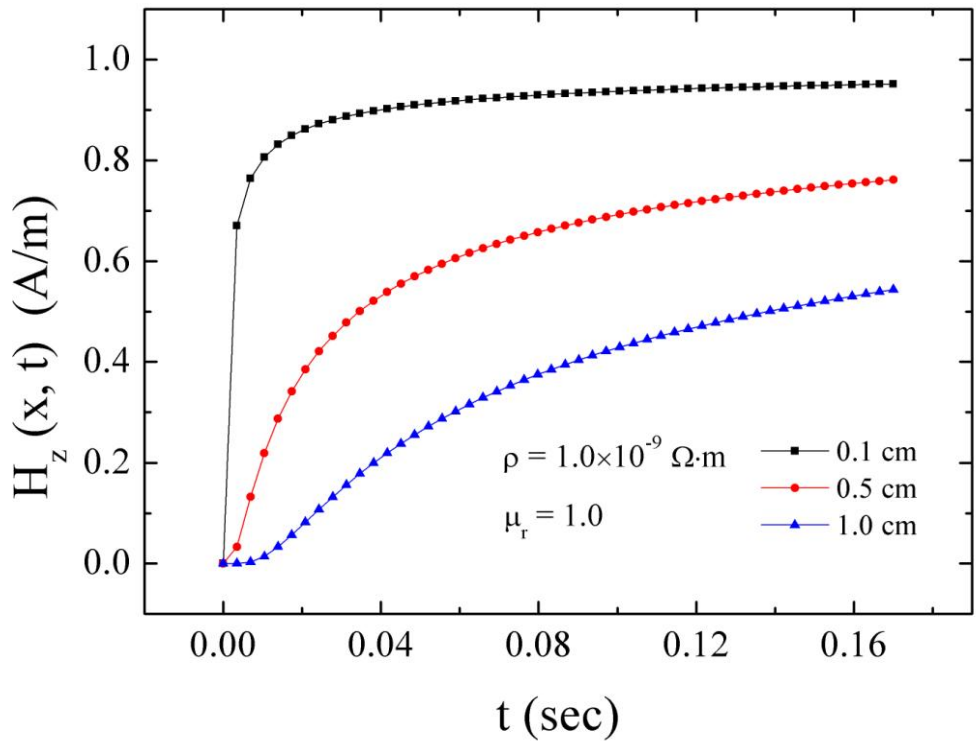

**Fig. 5:** The diffusion process at different positions $x$

2.2.2.1.2  Transient linear field $H_z(t) = (H_0/t_0)*t$

An external linear ramp is applied — as typically done in an accelerator: it starts from zero and reaches the maximum field $H_0$ after the time interval $t_0$.

The solution is found as follows:

$$H_z(x,t) = \frac{H_0}{t_0} t \left[ \left(1 + 2\xi^2\right) \operatorname{erfc} \xi - \frac{2}{\sqrt{\pi}} \xi e^{-\xi^2} \right] . \tag{20}$$

It can be split into three parts, of which the first two form the stationary part:

$$H_z(x,t) = \frac{H_0}{t_0} \cdot t - \tau_d(x) \frac{H_0}{t_0} + H_z^t(x,t) . \tag{21}$$

with

$$\tau_d(x) = \frac{x^2}{2\kappa} .$$

The first part is the applied external field. Owing to the eddy currents, the internal field lags behind during the ramp (second part) after the transient part (part 3) has vanished. The same is true at the end of the ramp. The situation is visualized in Fig. 6.

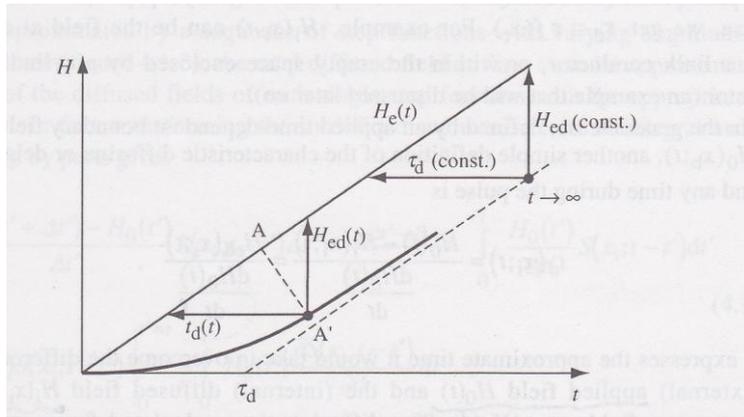

**Fig. 6:** Sketch of the time behaviour of the half-space field after the application of a linear ramp

Figure 7 visualizes $H_z(x, t)$ as a function of time for different positions $x$.

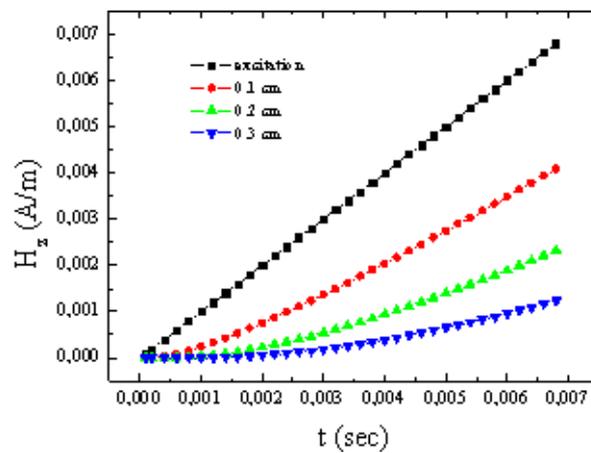

**Fig. 7:** Field diffusion as a function of time at different positions $x$

### 2.2.2.1.3 Transient sinusoidal field $H_z(t) = H_0 * \sin(\omega t)$

Solving the differential equations with this external field leads to the following stationary solution, after the transient part has disappeared:

$$H_z^S(x,t) = H_0 \cdot e^{-\frac{x}{\delta}} \cdot \sin(\omega t - \frac{x}{\delta}), \qquad (22)$$

with

$$\delta = \sqrt{\frac{2}{\omega \mu \sigma}}, \qquad (23)$$

$\delta$ is the well-known harmonic skin depth.

The field can penetrate the half-space only to the skin depth and has a phase shift relative to the externally applied field. Figure 8 shows the skin depth as a function of frequency and the product of conductivity and permeability (material property) [Knoepfel Fig. 4.2–5]. For example, 1 Hz-operation gives (for iron with $\mu_r = 100$) a skin depth of 20 mm (just connecting the corresponding points at the left and right scale). This diagram allows the magnet designer to estimate the lamination thickness of a laminated magnet, requiring that the skin depth $\delta$ be large compared to the lamination thickness. A lamination thickness of typically 0.5–1.0 mm seems to be adequate for the example mentioned.

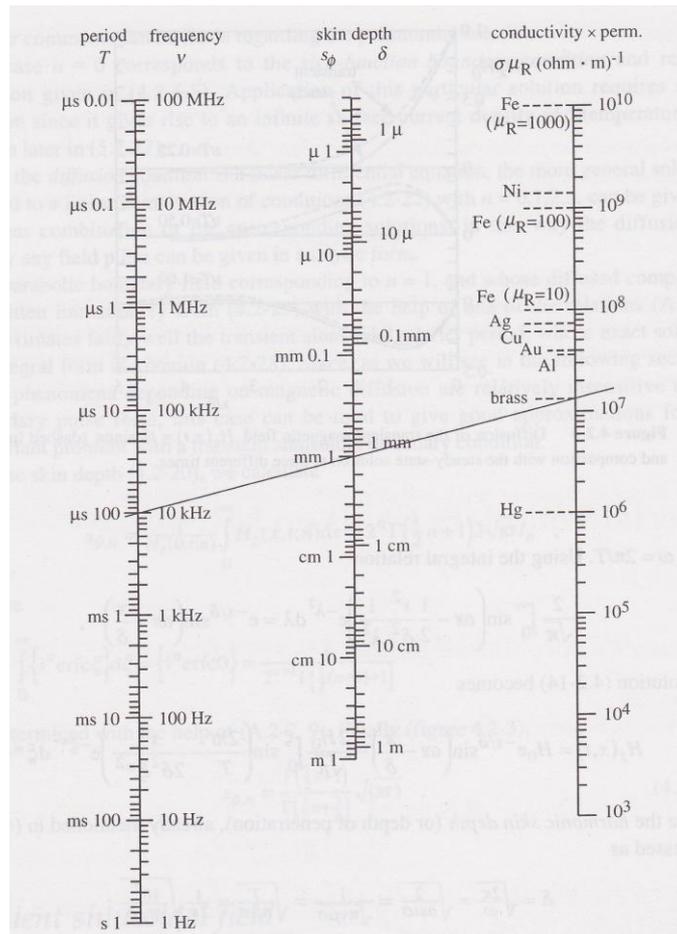

**Fig. 8:** Skin depth as function of frequency, conductivity and permeability [Knoepfel Fig. 4.2–5]

*2.2.2.2 Slab conductor*

The slab geometry is shown in Fig. 9. Only in the *x* direction does it have a limited dimension 2d.

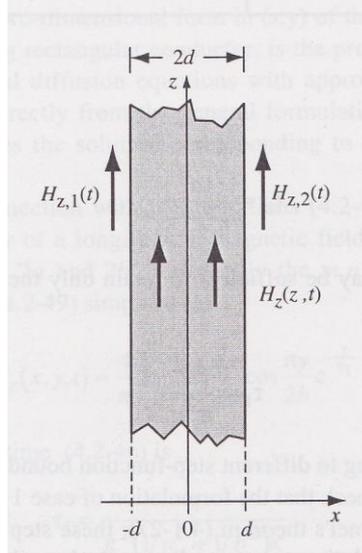

**Fig. 9:** Slab geometry (1-dimensional)

We consider step function excitation only, i.e., the following boundary conditions:

$$H_z(\pm d, t) = 0 \qquad t < 0$$

$$H_z(\pm d, t) = H_0 \qquad t \geq 0$$

$$H_z(x, 0) = 0 \qquad -d < x < +d$$

The solution of the magnetic diffusion differential equation yields

$$H_z(x,t) = H_0 \left[ 1 - 4 \sum_{n=1}^{\infty} \frac{\cos \frac{n\pi x}{2d}}{n\pi (-1)^{\frac{n-1}{2}}} e^{-\frac{t}{\tau_n}} \right], \qquad (24)$$

*n* odd, with

$$\tau_n = \frac{4}{n^2 \pi^2} \cdot \frac{d}{\kappa} \quad \text{and} \quad \kappa = \frac{1}{\sigma \cdot \mu}.$$

The second term in Eq. (24) is the transient term. It is the sum over different *n* including a whole spectrum of time constants $\tau_n$. Please note that the slab dimension appears as the square in the equation of the time constant. Doubling the lamination thickness leads to a factor of 4 in the time constant and, as we shall see later, also in the ohmic losses. One may call it the 'relevant distance' of the eddy currents. It is obvious that the time constants are inversely proportional to the diffusivity κ: the smaller the diffusivity (high conductivity, high permeability), the longer the time constants.

The longest possible time constant [for example, for a 1 mm thick low-carbon iron ($\mu_r$ = 1000, $\sigma = 10^7$ (Ohm*m)$^{-1}$) lamination] is less than 1 millisecond.

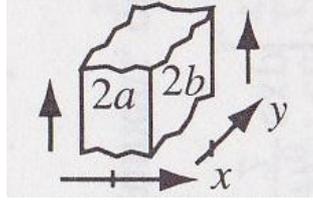

**Fig. 10:** Slab geometry (2-dimensional)

In a similar calculation the solution for a 2D slab (Fig. 10) can be found [Knoepfel]:

$$H_z(x,y,t) = H_0 \left[ 1 - 4 \sum_{n=1}^{\infty} \sum_{m=1}^{\infty} \frac{\cos\frac{n\pi x}{2a} \cos\frac{m\pi y}{2b}}{\pi^2 \cdot f(n,m)} e^{-\frac{t}{\tau_{n,m}}} \right] \quad (25)$$

*n,m* odd

$$\tau(n,m) = 4 / \left[ \pi^2 \kappa \left( \frac{n^2}{a^2} + \frac{m^2}{b^2} \right) \right]$$

$$f(n,m) = \frac{n*m*(-1)^{\frac{n+m}{2}}}{4}$$

### 2.2.2.3  *Iron-dominated C-dipole*

Of course a magnet builder is interested in how a real magnet behaves.

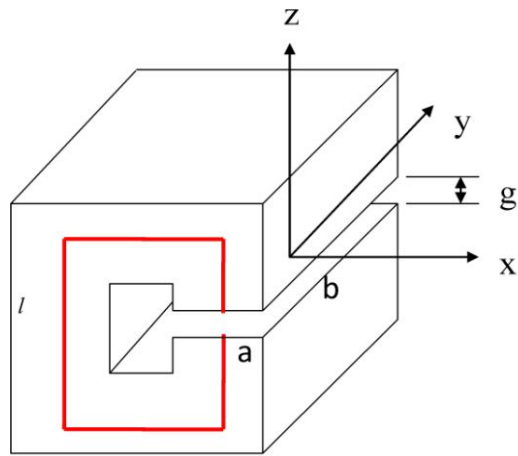

**Fig. 11:** C-shaped dipole

In a CERN internal note [3] the magnetic diffusion problem of a C-type dipole (Fig. 11) was handled. Here $l$ is the average iron path length, $g$ the gap height. Typically the ratio $l/g$ is about 10 ($l >> g$) and $a$ and $b$ correspond to the dimensions of the 2D slab in the previous subsection.

Starting from the magnetic diffusion equation, applied to a slab with a gap, they found a special differential equation for the field in the gap of the C-dipole:

$$\frac{\partial^2 H}{\partial x^2} + \frac{\partial^2 H}{\partial y^2} = \sigma \mu_0 \frac{l}{g + \left(l/\mu_r\right)} \frac{\partial H}{\partial t} \ . \tag{26}$$

It is easy to verify that for $g = 0$ we get the standard diffusion equation. For field levels away from saturation ($l/g << \mu_r$) we obtain

$$\frac{\partial^2 H}{\partial x^2} + \frac{\partial^2 H}{\partial y^2} = \sigma \mu_0 \frac{l}{g} \frac{\partial H}{\partial t} = \frac{1}{\kappa_1} \frac{\partial H}{\partial t} \qquad \kappa_1 = \frac{1}{\mu_0 \sigma * l/g} \ . \tag{27}$$

Here we defined a new diffusivity $\kappa_1$ that is not determined by the relative permeability, but by the dimensions $l$ and $g$ of the magnet. The magnetic energy is mainly stored in the gap of the magnet!

With this new diffusion constant all the previously mentioned solutions can be used. For the mentioned typical value of $l/g = 10$ we get a smaller value compared to the slab with $\mu_r$ between 100 and 1000 and consequently shorter time constants.

### 2.2.3 *Numerical solutions (introduction of numerical codes)*

Table 1 compares analytical with numerical methods.

**Table 1**: Comparison of analytical and numerical methods

|  | **Advantages** | **Disadvantages** |
|---|---|---|
| **Analytical methods** | • physical understanding | • simple geometry (mainly1D/ 2D) <br> • homogeneous, isotropic and linear materials <br> • simple excitation |
| **Numerical methods** | • complex geometry (3D) <br> • inhomogeneous, anisotropic and nonlinear materials <br> • complex excitation | • long computing times |

Analytical methods deliver a good physical understanding and the formulas allow a good estimate of the effects under simple assumptions, but for better results in complex situations one needs numerical calculations.

The most common way to define eddy currents in conducting media is based on solution equations for vector potential $A$ and scalar electric potential $V$ [4]:

$$\nabla \times \frac{1}{\mu} \nabla \times \vec{A} = -\sigma \frac{\partial \vec{A}}{\partial t} - \sigma V \quad , \tag{28}$$

$$\nabla \cdot \sigma \nabla V + \nabla \cdot \sigma \frac{\partial \vec{A}}{\partial t} = 0 \quad , \tag{29}$$

with appropriate boundary and conditions. The current density vector may be found as:

$$\vec{J} = -\sigma \frac{\partial \vec{A}}{\partial t} - \sigma \nabla V \quad . \tag{30}$$

Some codes use the current vector potential *T*, defined by

$$j = \sigma \cdot \nabla \times T \quad . \tag{31}$$

The following listing gives an overview of widely used commercially available numerical codes for the calculation of eddy currents in magnets.

- Opera© (Vector Fields Software, Cobham Techn. Services, Oxford) *www.vectorfields.com*
    o Finite Element Method (FEM)
    o Opera 2d,AC and TR, Opera 3d, ELEKTRA© , (TEMPO-thermal and stress-analysis)
- ROXIE© (Routine for the Optimization of Magnet X-Sections, Inverse Field Calculation and Coil End Design) (S. Russenschuck, CERN) https://espace.cern.ch/roxie/default.aspx,
    o BEM/FEM
    o Optimization of cos *nθ*-magnets, coil coupling currents only
- ANSYS© (ANSYS Inc.) *www.ansys.com*
    o Finite Element Method (FEM)
    o Direct and indirect coupled analysis (Multiphysics)
        ▪ eddy current → heat → rising temperature → change resistivity → change eddy current (very important, especially for sc magnets)

## 2.3 Direct solutions of Maxwell equations (small perturbation)

In Section 2.2 we looked at the eddy current effects from the perspective of field diffusion. Owing to eddy currents the magnetic field lags behind (Lenz's rule) and slowly diffuses into the conducting material. The magnetic diffusivity was defined. The field was calculated analytically by solving the diffusion equation. Formulas for the penetration depth and the time constants were given for some simple cases. Numerical codes were introduced which calculate field, current density, and losses in more complex cases. Basically, no special approximations are necessary.

In this section we want to use a second approach by solving the Maxwell equations directly. The eddy currents will be considered as a 'small perturbation'. Some people call the method 'quick and dirty', because it calculates the eddy currents from the Maxwell equations, not taking into account that these eddy currents also create a magnetic field which affects the current distribution as well. But under the assumption of a 'small perturbation', these are second-order effects.

Typically we consider a case as 'small perturbation', if the conducting material is 'magnetically thin', i.e., the relevant dimension is small compared to the penetration depth. Taking into account the formula for the skin depth

$$\delta = \sqrt{\frac{2}{\omega\mu\sigma}} \ , \tag{23}$$

we see that this is true for low frequencies (small ramp rates), low conductivity and low permeability.

This approach will be demonstrated with some examples.

### 2.3.1   *Eddy currents in a rectangular, long, thin plate*

Figure 12 shows the geometry.

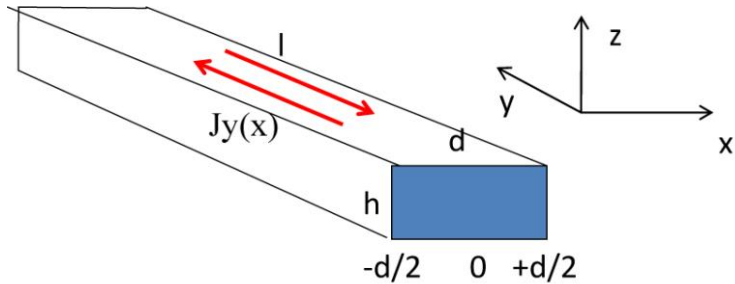

**Fig. 12:** Thin, long plate

The following conditions must be fulfilled:

- Field vertical to the plate, constant ramp rate $\dot{B}_z$
- Plate magnetically thin: $d \ll$ penetration depth $\delta$
- Plate geometrically thin: $h \ll d$
- Plate long enough to neglect the ends: $d, h \ll l$
- Steady-state conditions: $t \gg \tau_d$

From the integral form of Faraday's law one can calculate the eddy current distribution, neglecting the ends:

$$j_y(x) = -\frac{\dot{B}_z}{\rho} \cdot x \ . \tag{32}$$

From Ampere's law (1), one then obtains the field distribution produced by the eddy currents

$$H_z^{eddy}(x) = \frac{\dot{B}_z}{2\rho}\left(x^2 - \frac{d^2}{4}\right) , \tag{33}$$

and finally, the loss by integration:

$$dP = \rho \frac{l}{A} \cdot (j_y(x) \cdot A)^2 = \rho \cdot l \cdot j_y^2(x) \cdot h \cdot dx \qquad (34)$$

$$P = 2 \int_0^{d/2} dP = \frac{lh}{12} \frac{d^3}{\rho} \dot{B}_z^2 \quad or \quad P/volume = \frac{1}{12} \frac{d^2}{\rho} \dot{B}_z^2 . \qquad (35)$$

Please note:

- The relevant width $d$ enters the formula as squared. The power loss per volume depends only on the square of the relevant dimension $d$, the resistivity ρ, and of course on the square of the ramp rate.
- The eddy currents produce a negative dipole and a sextupole (for example on the top and bottom of a rectangular beam pipe).
- Because of the geometrical restriction, the loss formula is the same as for a thin slab (thin lamination).
- The resistivity ρ is a material property (Fig. 13). For copper it is more than two orders of magnitude lower than for iron or stainless steel. Therefore beware of eddy currents in copper! The resistivity depends on temperature (Fig. 14). At 4 K it is up to one order of magnitude lower than at room temperature. Therefore, the eddy current losses in a superconducting magnet are much higher than in a room-temperature magnet (besides the fact that they occur at 4 K, where the Carnot efficiency is very low).

| Resistivity ρ (Ohm*m) | 300K | 4K |
|---|---|---|
| LC steel (3% Silicon) | $590 \times 10^{-9}$ | $440 \times 10^{-9}$ |
| Stainless steel | $720 \times 10^{-9}$ | $490 \times 10^{-9}$ |
| Copper | $17.4 \times 10^{-9}$ | $0.156 \times 10^{-9}$ |

**Fig. 13:** Typical resistivity for different materials

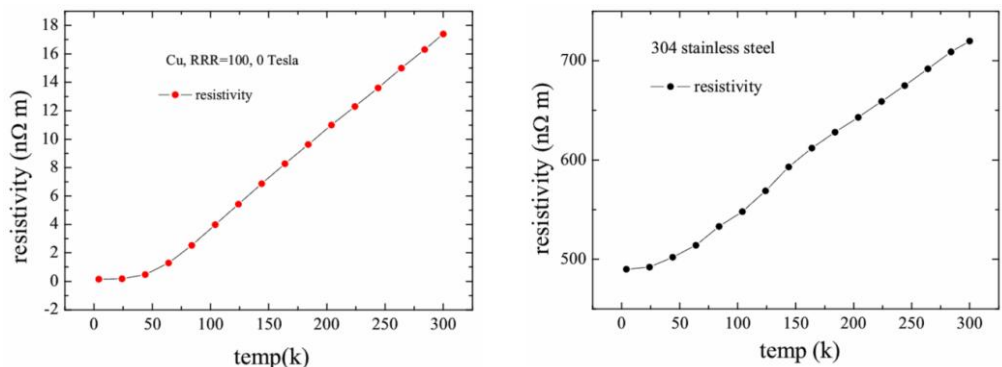

**Fig. 14:** Resistivity as a function of temperature for copper and stainless steel

Figures 15 and 16 compare a 'small perturbation' calculation with a numerical FEM solution. One can clearly see that for the stainless-steel case the results agree, but for copper (with its very high electrical conductivity) the condition 'magnetically thin' is no longer fulfilled and the results are completely different.

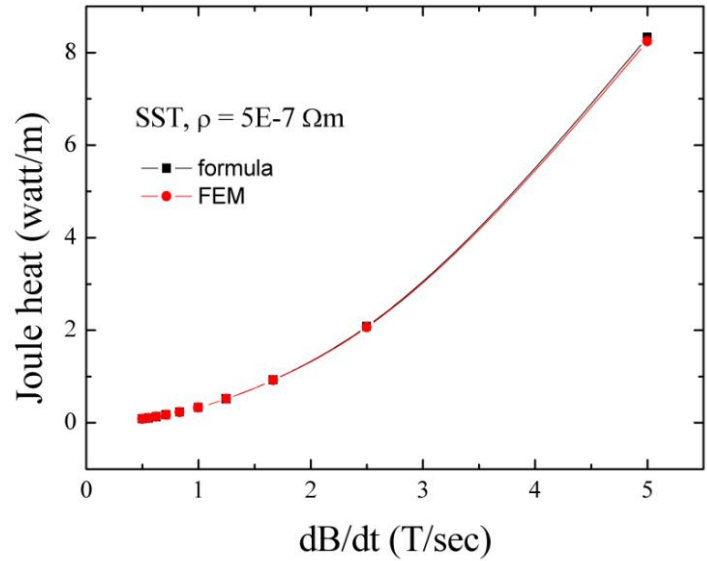

**Fig. 15:** Joule heating in thin stainless-steel plate, the numerical result is well reproduced by the approximation

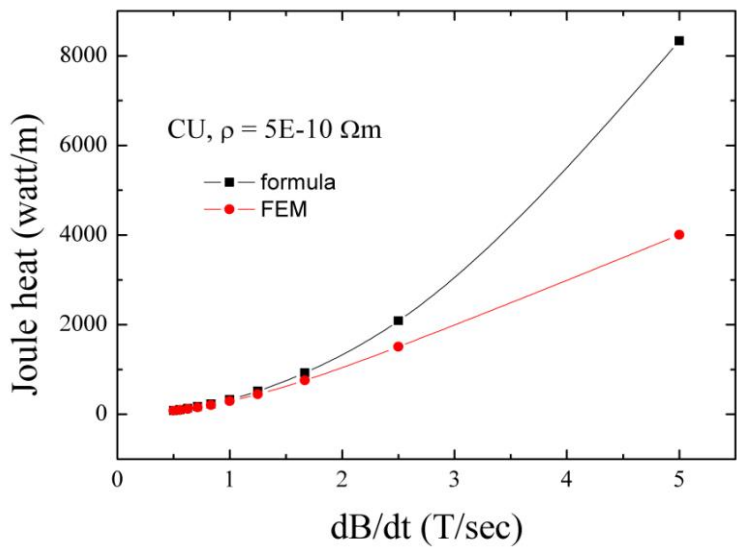

**Fig. 16:** Joule heating in a copper plate, the 'small perturbation' condition is no longer fulfilled due to the high conductivity of the copper

### *2.3.2 Eddy currents in a long, thin cylinder (beam pipe)*

This example handles a thin cylinder as used in many accelerator magnets. The cylinder with radius $r$, wall thickness $d$ and length $l$ must be magnetically thin (i.e., $r \ll$ skin depth), geometrically thin $d \ll r$ and long ($d, r \ll l$), to ignore the ends and therefore be handled as a 2D problem.

Then, we obtain (in the same way as in 2.3.1) the current density and the power loss

$$j = \frac{r\cos\theta}{\rho}\dot{B} \ . \tag{36}$$

$$P = \frac{r^3}{\rho}\dot{B}^2 \pi dl \quad \text{or} \quad P/V = \frac{r^2}{2\rho}\dot{B}^2 \ . \tag{37}$$

Please note:

- The eddy current distribution is proportional to $\cos\theta$, which leads to a pure dipole.
- Again, the loss per volume shows the same dependence as for the thin plate.

### 2.3.3 *Eddy currents in a round thin plate or disk*

In a thin plate or disk with radius $r$ and thickness $d$ (with $r \gg d$) we obtain, with the same computation steps

$$P/V = \frac{r^2}{8\rho}\dot{B}^2 \ . \tag{38}$$

## 3 Eddy currents in accelerator magnets

In this section we consider real accelerator magnets and investigate the role of eddy currents in these magnets [5–7].

Eddy currents can occur in all conductive elements which see a change in magnetic flux, especially in:

- iron yokes (low carbon iron)
- mechanical structures (mostly low carbon iron or stainless steel)
- coils (copper, superconductor)
- beam pipes (stainless steel)

The main classification of these magnets is:

- coil-dominated magnets ( cos $n\theta$ current distribution or intersecting ellipses)
- iron-dominated magnets

Figure 17 shows some typical examples.

Both types of magnet can be equipped with resistive or superconducting coils. Nevertheless iron-dominated magnets are generally used for fields below 2 T due to iron saturation and are therefore mostly resistive, while coil-dominated magnets use superconducting coils for higher field strength.

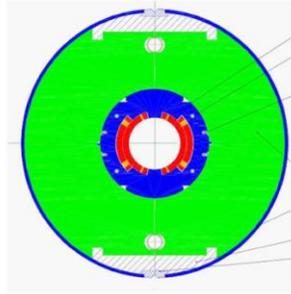 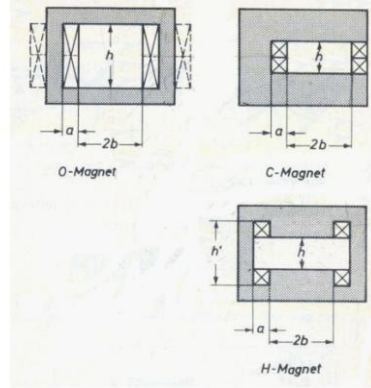

**Fig. 17:** Some magnet types

## 3.1 Eddy currents in the yoke

Accelerator magnets can have either solid or laminated yokes. Since most magnets which are subjected to a time varying field have laminated yokes we shall treat only those.

### *3.1.1 2D effects within a long magnet*

The first questions to answer are the choice of the lamination thickness and of the iron properties.

#### *3.1.1.1 Lamination thickness d*

In order to keep the influence of eddy currents small we need to keep

- the relevant geometrical dimensions (here the lamination thickness) small compared to the penetration depth (refer to 2.2.2.1.3). Using Eq. (23) or Fig. 8 one can find the penetration depth $\delta$:

$$\delta = \sqrt{\frac{2}{\omega\mu\sigma}} \ ; \tag{23}$$

- the longest time constant (refer to Eq. (24) for a one-dimensional slab or Eq. (27) for a C-shaped dipole) small compared to the ramping time or cycle time of the application:

$$\tau_n = \frac{1}{n^2\pi^2} \cdot \frac{d^2}{\kappa} \ ;$$

please note that in this formula the thickness of the lamination is *d*, not 2*d*!

with

$$\kappa = \frac{1}{\sigma \cdot \mu} \quad \text{for a slab},$$

$$\kappa = \frac{1}{\sigma \cdot \mu_0 * l/g} \quad \text{for a C-shaped dipole}.$$

In addition the power loss/volume in the laminations due to the eddy currents should be calculated [Eq. (35)]:

$$P/volume = \frac{1}{12}\frac{d^2}{\rho}\dot{B}_z^2.$$

The losses in the laminations should be small or at least comparable to other losses in the magnet.

Please note that in this formula also the lamination thickness is $d$, not $2d$!

*Conclusion*: In order to limit the influence of the eddy currents, we build magnets with carefully insulated laminations of high resistivity iron. Obviously, one has two parameters to play with: the lamination thickness and the resistivity.

### 3.1.1.2  Choice of the iron

The following three pictures were taken from Ref. [8].

Silicon increases the resistivity of low carbon iron (Fig. 18).

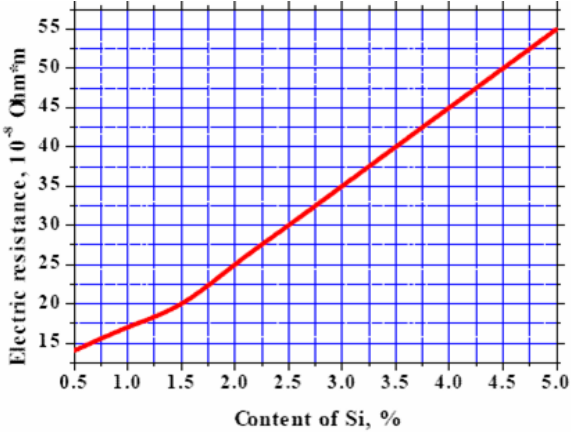

**Fig. 18:** Resistivity of iron as a function of the content of silicon

Fortunately, silicon also reduces the coercivity, therefore reducing the hysteresis losses in the laminated yoke (Fig. 19).

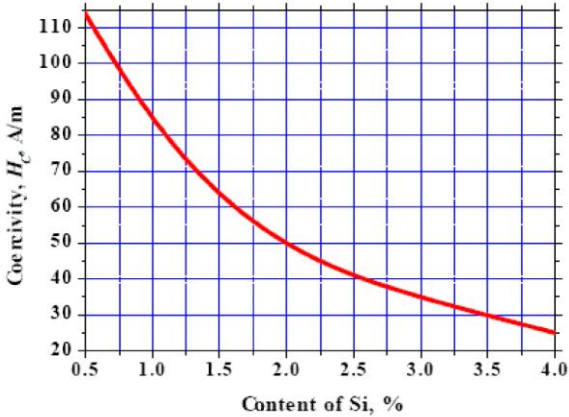

**Fig. 19:** Coercivity of iron as a function of the content of silicon

Unfortunately, the saturation magnetization decreases with the silicon content, defining a limit of 3.5% silicon content (

Fig. **20**).

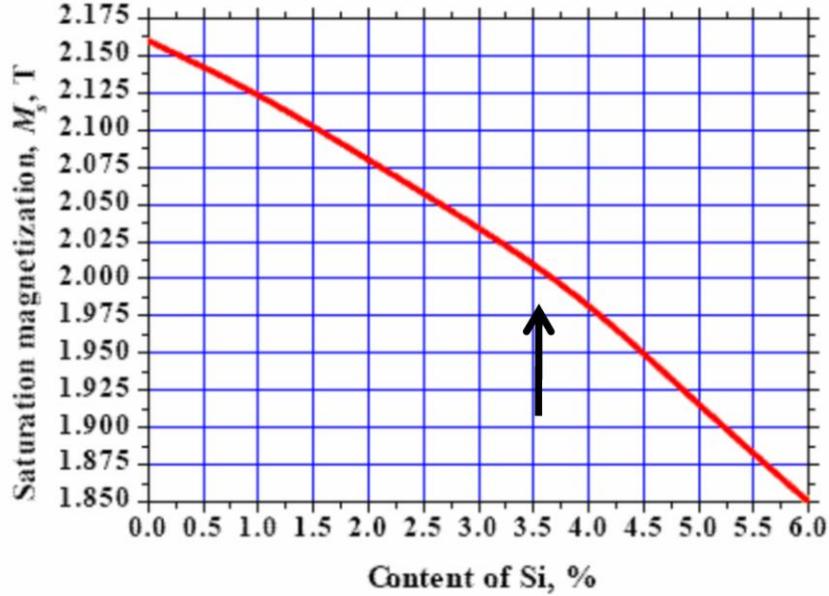

**Fig. 20:** Saturation magnetization as a function of the content of silicon

There is a practical limit on lamination thickness too: steel sheets are available on the market down to 0.3 mm (transformers) thickness. In addition, thinner laminations reduce the packing factor and increase the labour costs.

Please note: steel suppliers typically give total iron losses at 50 Hz. They have three parts:

- eddy losses, scaling with the square of the frequency;
- hysteresis losses, scaling linearly with the frequency; and the so-called
- anomalous losses, scaling with the power of 1.5 of the frequency.

Eddy current losses are given by

$$P\left[W/m^3\right] = \frac{\pi^2 v^2}{6\rho} B_p^2 d^2 \tag{39}$$

$v$ - frequency (Hz)

$d$ - lamination thickness (m)

$\rho$ - resistivity (Ohm*m)

$B_p$ - induction amplitude (T)

Hysteresis losses can be measured with a permeameter. The rest are anomalous losses. By appropriate scaling with the frequency, you may find the losses for your application [9].

*3.1.2    3D effects*

*3.1.2.1   Anisotropy of a laminated yoke*

3.1.2.1.1   Packing factor

We define a packing factor $f_p$

$f_p = W_i / (W_i + W_a)$     with

- $W_i$ thickness of a single lamination (without insulation),
- $W_a$ thickness of insulation.

In reality, the packing factor is slightly lower for mechanical reasons. Typical packing factors are between 0.95 and 0.98.

3.1.2.1.2   Conductivity

Since the laminations are insulated, the conductivity is highly anisotropic:

- perpendicular to the lamination: $\sigma_z = 0$
- within the lamination:   $\sigma_{xy} \neq 0$.

3.1.2.1.3   Relative iron permeability $\underline{\underline{\mu_r}}$

Generally, the relative permeability is a tensor. Each element $\mu_{ij}(x,y,z;\vec{H})$ depends on position and field strength. We call the permeability homogeneous if it does not depend on position, and linear if it does not depend on $\vec{H}$. It is isotropic if all diagonal elements are equal to $\mu_r$ and the non-diagonal elements are zero.

In the case of a laminated magnet the permeability is basically homogeneous, but it is strongly non-linear, especially at low field, due to coercivity, and at high field, due to saturation.

It is highly anisotropic, since the diagonal elements differ from each other. The permeabilities $\mu_{xx}$ or $\mu_{yy}$ within the lamination are different from the permeability $\mu_{zz}$ perpendicular to the lamination. The permeabilities can be calculated using the continuity equations:

For magnetic flux tangential to the laminations [10] :

$$\mu_{xx} = \mu_{yy} = f_p * (\mu_r(\vec{H}) - 1) + 1 \approx f_p * \mu_r(\vec{H}) \ . \tag{40}$$

The flux is reduced by the packing factor.

For magnetic flux normal to the laminations [10]:

$$\mu_{zz} = \mu_r(\vec{H}) / (\mu_r(\vec{H}) - f_p * (\mu_r(\vec{H}) - 1)) \ , \tag{41}$$

and for $\mu_r \gg 1$

$$\mu_{zz} = 1/(1-f_p) \ . \tag{42}$$

In this case, the permeability is drastically reduced to typical values between 20 and 50 at low fields and will further decrease for saturation.

Figures 21 and 22 show the permeabilities as a function of field [11].

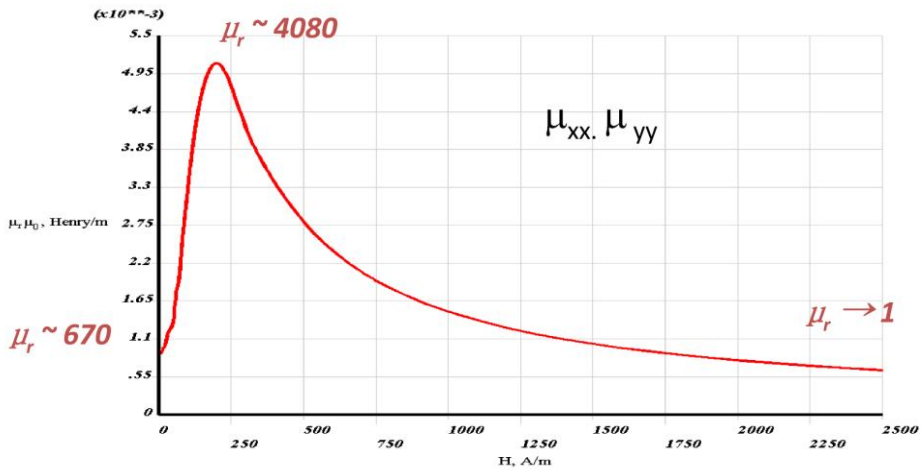

**Fig. 21:** Permeability $\mu_{xx}$ and $\mu_{yy}$ (in the plane of the lamination)

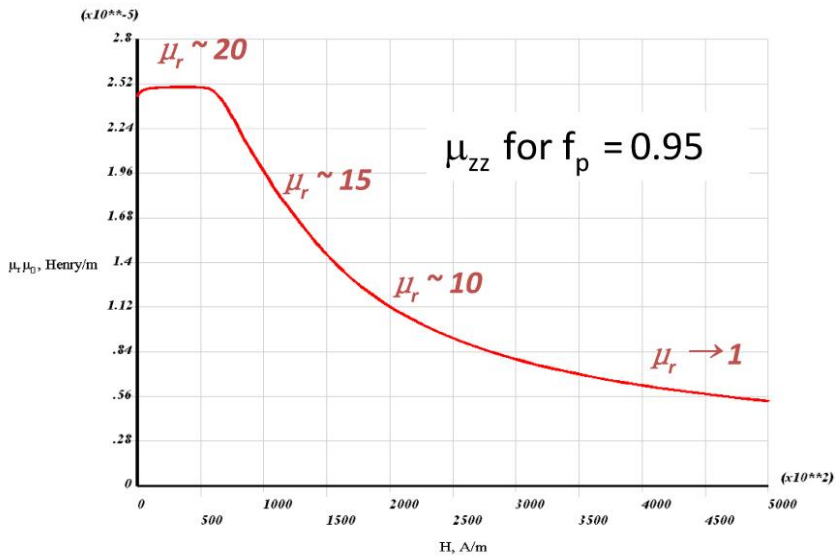

**Fig. 22:** Longitudinal permeability $\mu_{zz}$

One sees clearly that for low fields the permeability within the lamination is much higher than the longitudinal permeability because the magnetic reluctance is much higher in this direction. However, at high fields due to iron saturation the permeabilities become comparable. As a consequence the distribution of magnetic flux changes along a magnetic ramp from low to high field — just a static phenomenon! Near the magnet ends the longitudinal flux penetrates the iron at high field more than at low field. Since this flux change is responsible for the eddy current loss in the ends, the losses will increase at high fields.

*3.1.2.2 Eddy currents due to field components $B_z$ perpendicular to the laminations*

An ideal (infinitely long in the *z*-direction) magnet can be considered as a 2D structure. It has no *z*-component of the field and its properties do not vary with the longitudinal coordinate *z*. As was shown in the last paragraph we optimize the magnet as far as eddy currents are concerned by choosing

- the appropriate lamination thickness *d* (practical limit 0.3 mm),
- low steel conductivity,
- low coercivity (in order to reduce the hysteresis losses).

The dimension relevant for eddy current in this case is the lamination thickness *d* (Fig. 23).

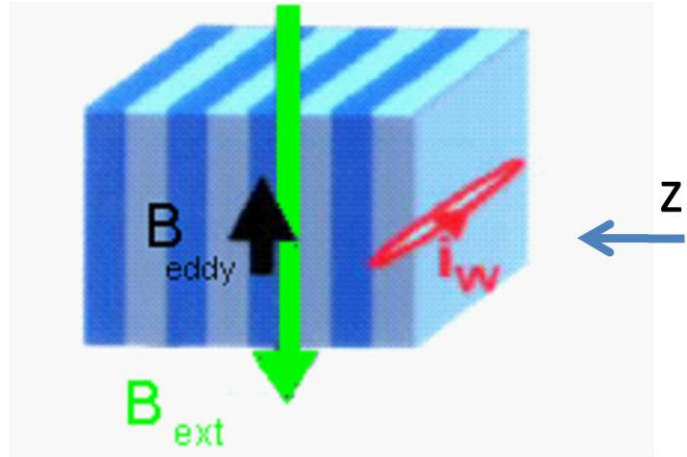

**Fig. 23:** Schematic view of a 2D-magnet (no $B_z$ component)

A real magnet has of course a limited length and its properties — such as the packing factor — do vary with *z*. Therefore longitudinal field components $B_z$ exist in

- yoke end regions,
- areas with varying packing factor etc. (Fig. 24).

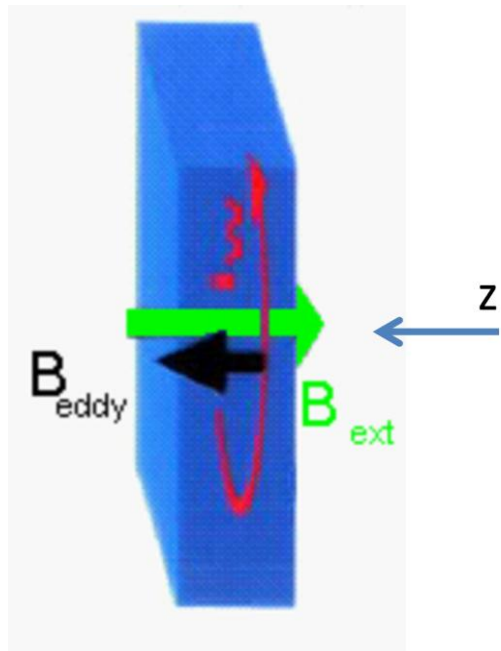

**Fig. 24:** Schematic view of a 3D-magnet ($B_z$ components exist)

In this case the relevant dimension of the eddy current pattern is not the lamination thickness anymore, but the *x-y* surface itself (or part of it). Consequently, time constants are larger and losses are not negligible.

3.1.2.2.1  Field variation due to eddy current effects in the end part of magnets

In the following paragraphs time delay and losses caused by eddy currents in the end part of real magnets will be demonstrated. The eddy currents are induced by a linear ramp of the excitation current or a triangular cycle. The corresponding field lag and time constant is a function of *z* and can be measured.

Figure 25 schematically shows the situation at the end of a dipole.

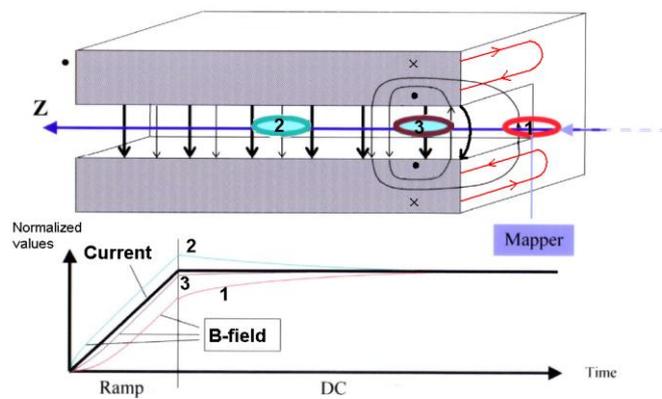

**Fig. 25:** Schematic view of the eddy current induced vertical field component (courtesy of F. Klos)

The *z* component of the main field induces eddy currents in the pole sheets, which produce their own field, superimposing the main field. In region 1 the vertical field component $B_y$ decreases (field lag), while in region 2 the vertical field component increases. In the intermediate region 3 we have a transition. Of course the effect becomes smaller towards the longitudinal magnet centre.

3.1.2.2.1.1  SIS 18 dipole

This phenomenon was tested experimentally on the 2.6 m long curved dipole of the synchrotron SIS 18 at GSI (Fig. 26).

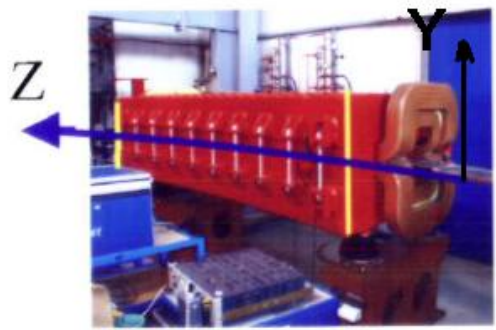

**Fig. 26:** The synchrotron dipole of SIS18 at GSI

The magnet was ramped up to 1.8 T in 1.1 seconds. The vertical field $B_y$ was measured with a Hall probe along the trajectory $z$. In Fig. 27 is shown the measured field lag/enhancement as a function of time ($t = 0$ corresponds to the end of the current ramp) and longitudinal position $z$ ($z = 0$ corresponds to the longitudinal magnet centre).

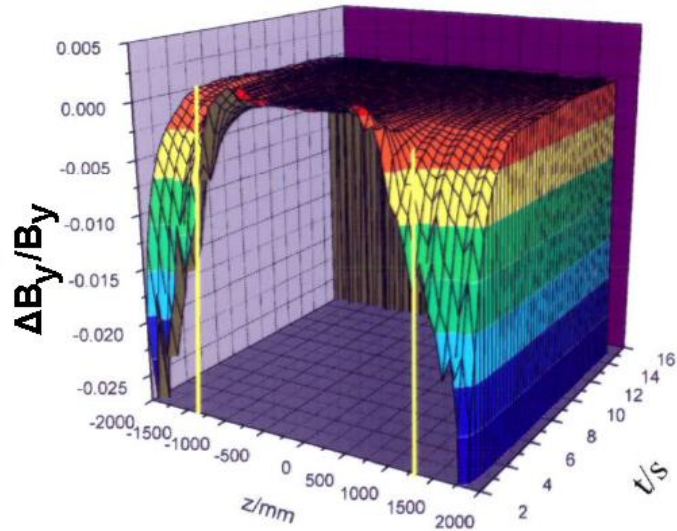

**Fig. 27:** Field lag/enhancement of the vertical field component due to eddy currents ($t = 0$ corresponds to the end of the ramp, longitudinal magnet centre at $z = 0$, yoke end at $z = \pm 1300$ mm)

One clearly recognizes the field lag outside, but also inside of the magnet. Since the magnet is ramped up to saturation the $B_z$ components exist far inside of the magnet (refer to 3.1.2.1.3).

The much smaller effect of field enhancement is observed only near the center of the magnet. An integral measurement with a search coil shows only an integral field lag (Fig. 30).

Calculations based on current vector potential verified these experimental results [12]. The current density of the pole surface as a function of $z$ at different times is shown in Fig. 28. One realizes that the eddy currents (induced by the $B_z$ components) penetrate into the magnet at higher excitation. That happens because the $B_z$ flux penetrates into the iron due to the anisotropy of the permeability, mentioned earlier.

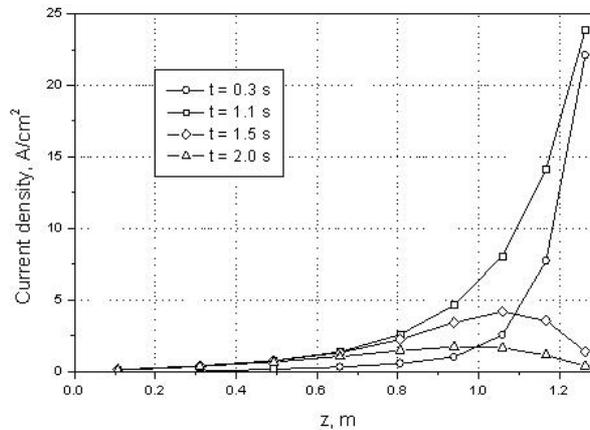

**Fig. 28:** Current density of the pole surface as a function of $z$ (ramp starts at $t = 0$ and ends after 1.1 seconds).

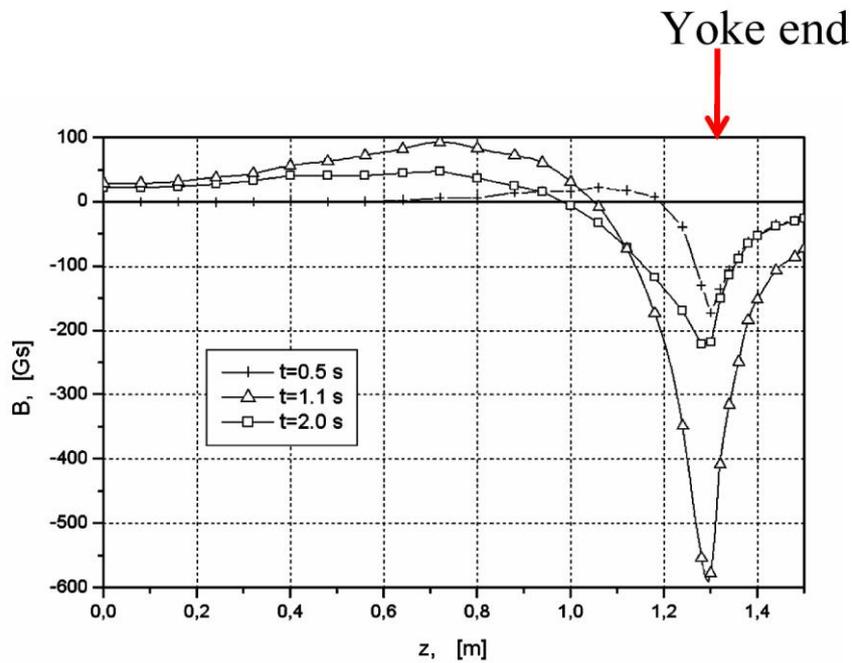

**Fig. 29:** $B_y$ field component, induced by the eddy currents

Figure 29 confirms the field lag/enhancement of $B_y$ and Fig. 30 compares the measured and calculated $B_y$ (induced by the eddy currents) integrated over the whole magnet length.

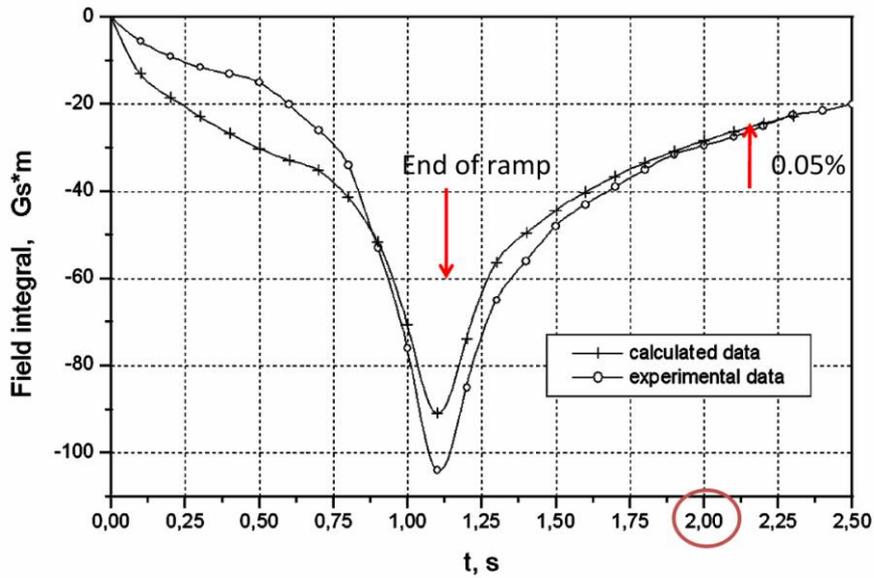

**Fig. 30:** Integral field induced by eddy currents (ramp 0–1.8T in 1.1 seconds)

The effect is huge: only 2 seconds after the end of the ramp the gap field reaches the required field quality of $5 \times 10^{-4}$ (total field integral 48000 Gs·m).

3.1.2.2.1.2    CNAO dipole

Field delay was also measured on a 0.44 m long scanner dipole of CNAO (0.3 mm lamination thickness). It was ramped with 500 T/s up to 0.3 T. The gap field $B_y$ was measured with a Hall probe at several longitudinal positions $z$. The field lag was fitted to an exponential function after the ramp stopped. The results are shown in Fig. 31 [13].

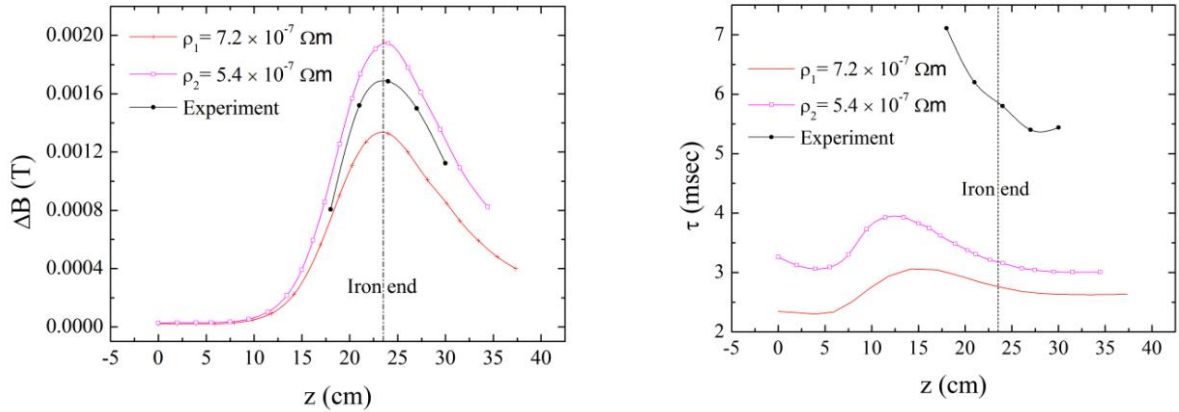

**Fig. 31:** Field lag and diffusion time constant as a function of the longitudinal coordinate $z$. ($z = 0$ corresponds to the magnet centre).

One can observe that

– A significant field lag occurs only near the magnet end; the eddy currents do not penetrate much, since the magnet is operated at low field (max. 0.3 T).

– The diffusion time constant near the ends is of the order of some milliseconds, much larger than the value calculated with Eqs. (24) and (27) for the centre of the magnet.

– The field lag is up to 0.7% of the nominal field (the value integrated over the whole magnet is of course smaller).

– The numerical values, calculated with ANSYS, agree reasonably well with the measured data.

3.1.2.2.1.3    SIS 100 dipole model

This type of dipole (2.72 m long, superferric, window-frame type) will be used as synchrotron dipole of the SIS 100 of the FAIR facility. It has a superconducting coil, forming the cold mass together with cold iron. Since it is a synchrotron magnet, correct tracking requires the time lag of the integral field to be known.

The field lag was calculated by ANSYS at four different longitudinal positions [14] as indicated in Fig. 32.

Outside of the magnet, we observe a large field lag and inside at positions 0.689 and 1.01 m a small field enhancement. At the position 1.17 m the initial field enhancement transforms into a field lag, since the eddy current centre moves inside the magnet. All time constants are of the order of 100 ms, smaller than the ramp time of 475 ms.

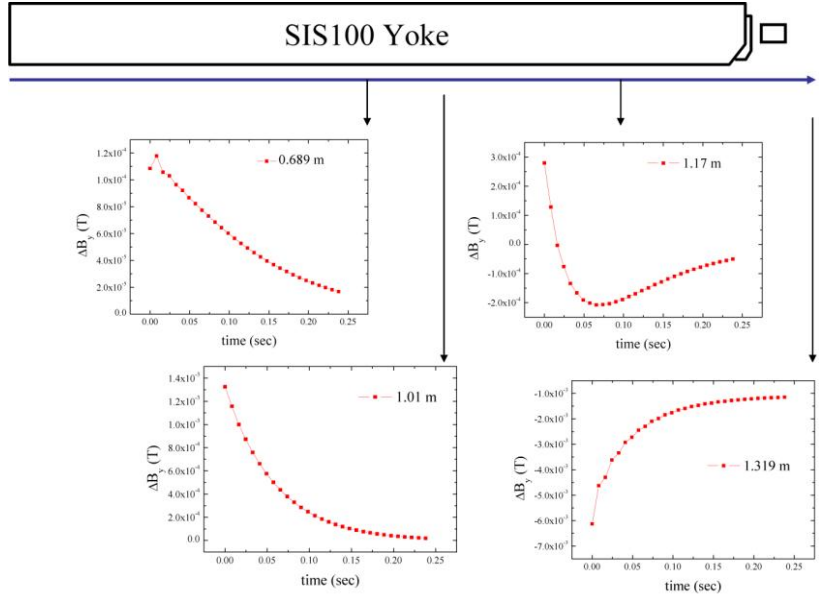

**Fig. 32:** Field lag/enhancement at several longitudinal positions ($t = 0$ corresponds to the end of a linear ramp up to 1.9 T, ramp rate 4 T/s, $x = 0$ corresponds to the centre, $x = 1.36$ m to the yoke end plate)

3.1.2.2.2  Resistive losses due to eddy current effects in the end parts of magnets

As mentioned before, the eddy current resistive losses are of special interest for superconducting cold iron magnets. Besides the hysteresis losses they form the biggest contribution to the cryogenic load.

3.1.2.2.2.1  SIS 100 dipole

Loss calculation results can be found in several publications [15], [16].

Figure 33 shows the calculated end-part losses during a triangular cycle with a period of 1 s [17], [18]. The current ramp rate is kept constant. Owing to the nonlinearity and anisotropy of the permeability (refer to 3.1.2.1.3) the losses are not constant during the ramp. They depend strongly on the packing factor: a large packing factor leads to a large longitudinal permeability, which reduces the magnetic reluctance and allows the flux to penetrate. Therefore the eddy current density penetrates further into the iron increasing the losses. A well stacked magnet may produce large end-part losses. The diffusion time constants and losses are investigated as a function of operating frequency and stacking factor in Ref. [19].

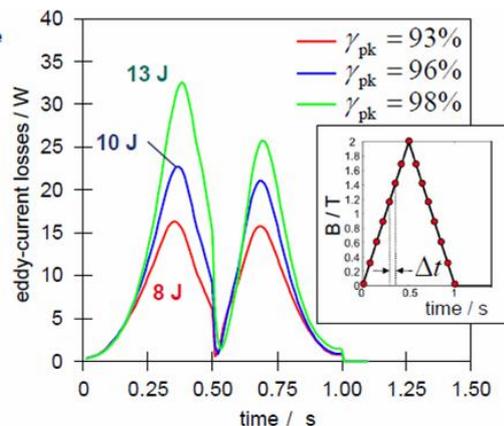

**Fig. 33:** End-part losses during a triangular cycle (parameter is the packing factor)

Figure 34 makes that even clearer: it shows the loss distribution as a function of the longitudinal coordinate. The variable parameter is the time during the cycle [11], [20]. The higher the field, the more the eddy current density penetrates into the iron.

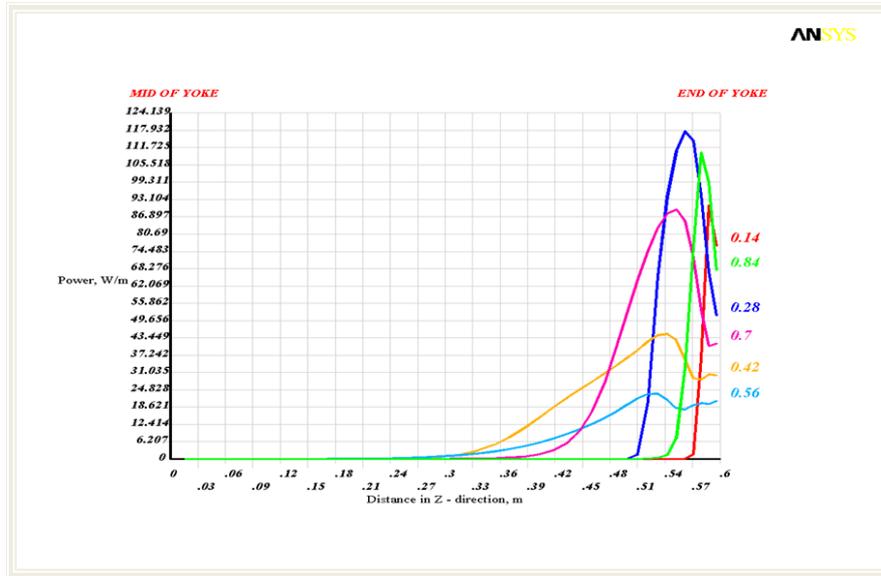

**Fig. 34:** Power loss as a function of longitudinal coordinate *z*, at different times

3.1.2.2.2.2    SIS 300 dipole

This dipole is shown schematically in Fig. 35. It is a superconducting cosθ dipole with coil, collar, low-carbon iron and stainless-steel laminations, replacing the iron at the magnet ends. The coil is longer than the iron in order to lower the high field point. Flux created by the end of the coil enters the iron with a large $B_z$ component which generates eddy currents in the iron lamination and in the stainless steel and collar laminations as well [21].

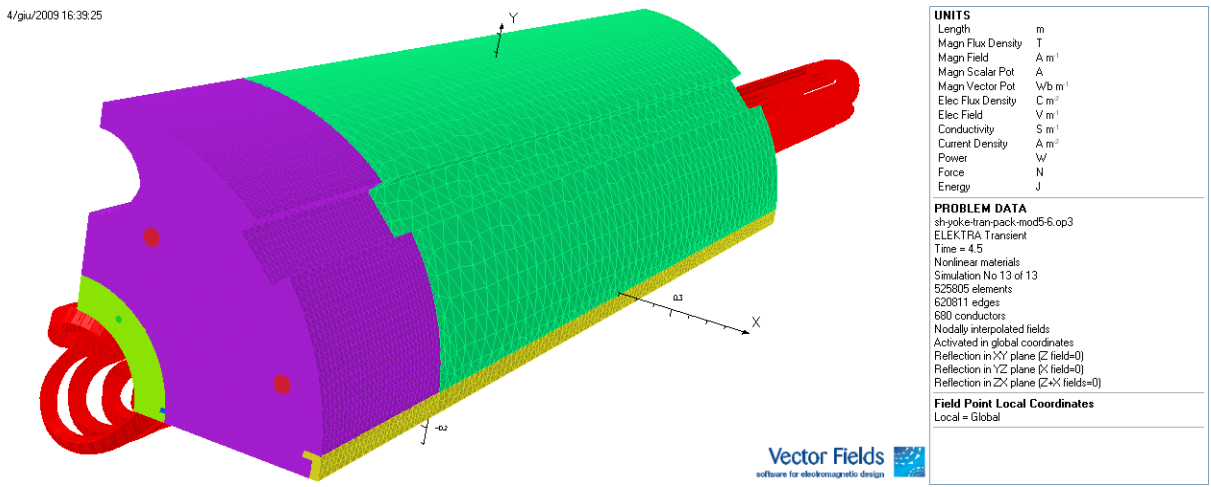

**Fig. 35:** Schematic view of the SIS 300 prototype dipole (coil, collar, low-carbon iron and stainless-steel ends)

The eddy current pattern was calculated using ELEKTRA$^©$. At 4.5 T, on account of saturation the eddy currents penetrate more in the iron and create there higher losses than at 1.5 T [9]. We observe here on a cosθ dipole the same effect which we have previously seen on the superferric SIS 100 dipole.

3.1.2.2.3   Field variation due to variation of the packing factor

Usually the packing factor of a laminated magnet varies slightly along the trajectory on account of mechanical imperfections. Stacked laminated dipoles with large bending angle have to be built out of several sectors. At the boundaries the packing factor is of course reduced, the DC field $B_y$ therefore drops and field components perpendicular to the lamination surface are created. Figure 36 shows this effect schematically. Figure 37 shows a plot of the DC field and the field lag after a magnet ramp along the trajectory. One can clearly see the coincidence of both effects at the five interfaces of the six magnet sectors [22].

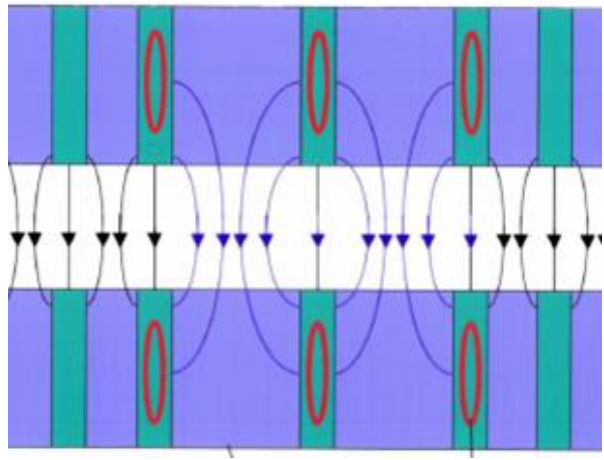

**Fig. 36:** Eddy currents at positions of low packing factor due to field components perpendicular to the lamination surface (courtesy of F. Klos)

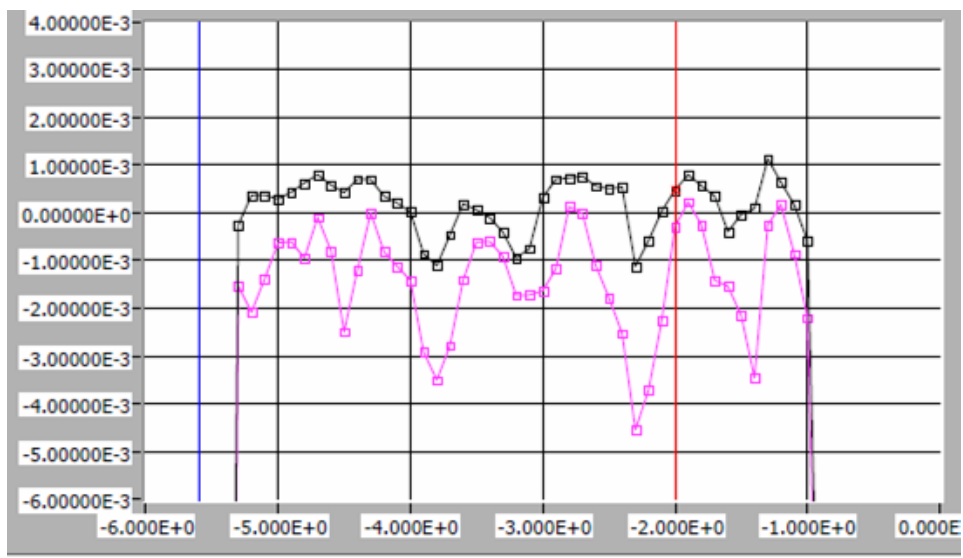

**Fig. 37:** Coincidence of DC field drop (dark dots) and field lag (light dots) due to eddy currents at the five sector interfaces (points with low packing factor)

## 3.2 Eddy currents in coils

### 3.2.1 *Eddy currents in resistive coils*

As mentioned in the introduction, eddy current effects in accelerator magnets are mostly unwanted. However, there is an example where the homogeneity of a synchrotron magnet benefits from the eddy current contribution to the field [23].

Figure 38 shows the cross-section of a typical 'hybrid' dipole. It is a mixture of an H-type dipole and a window-frame dipole. In order to reach the necessary ampere-windings for high field operation one has to place part of the coil between the poles. Such dipoles were built for the SPS at CERN and SIS18 at GSI.

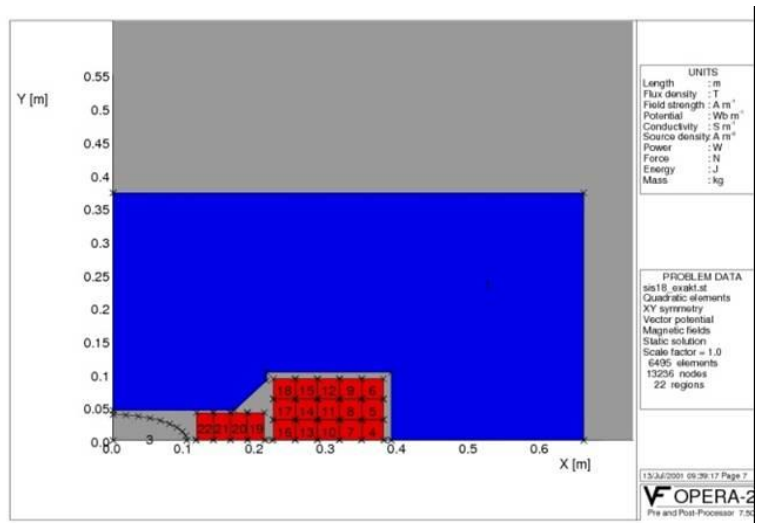

**Fig. 38:** Cross-section of a 'hybrid' dipole

Figure 39 demonstrates the effect. While the transport current (shown in conductor 1-5 with an 'x') is ramped up, eddy currents are induced in each of the massive copper conductor (here shown only in conductor 6), which (for symmetry reasons) introduce a positive sextupole-like component.

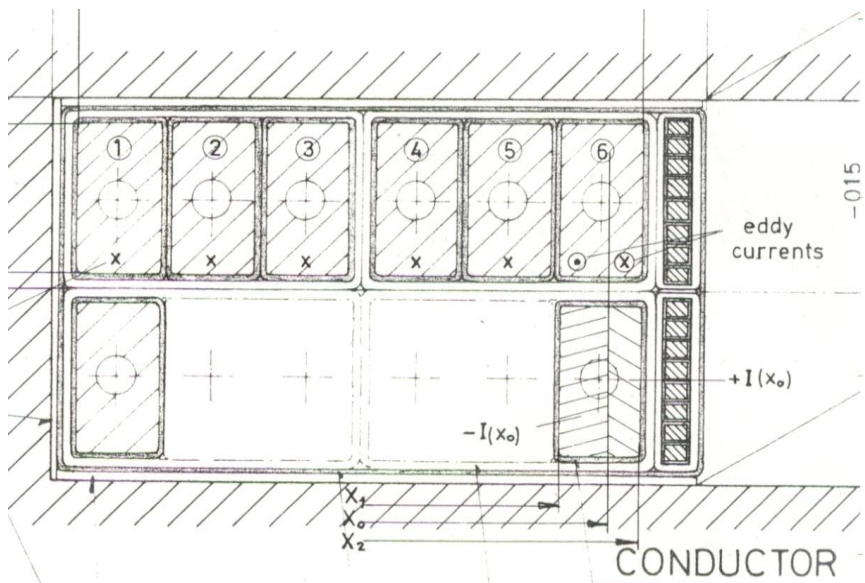

**Fig. 39:** Cross-section of a copper coil between the pole shoes of a 'hybrid' dipole

Figure 40 shows the measured field quality, in the DC mode (three currents) without and in the ramped mode (maximum current 380 A) with eddy currents. The eddy currents compensate the negative sextupolar component. For a ramp rate of 1000 A/s the field quality is almost ideal.

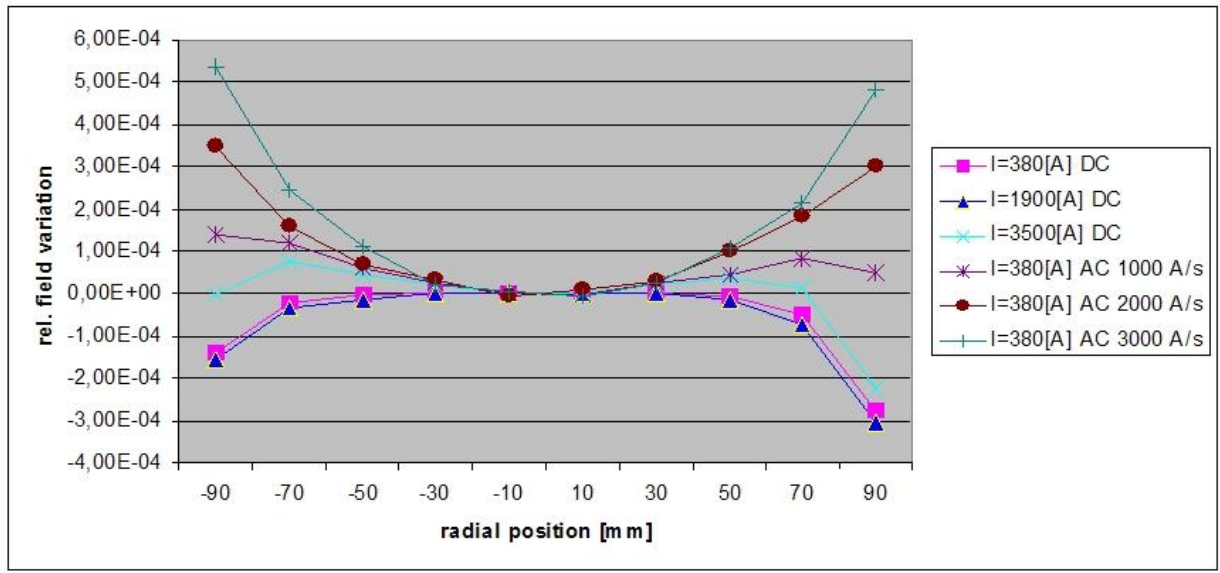

**Fig. 40:** Field quality in DC mode and at different ramp rates (Courtesy of F. Klos)

### 3.2.2  *Eddy currents in superconducting coils*

Coupling between strand filaments and coupling between the strands of a cable are (besides magnetization) the biggest source of losses and field quality distortion in a superconducting coil. Eddy currents in the copper matrix of the strand also play a role. These effects are covered extensively in the literature [Wilson, Verweij].

## 3.3  Eddy currents in mechanical structure

The mechanical structure of a magnet normally consists of metal parts to provide a stiff structure with a high Young modulus. Therefore, eddy currents are induced if these metallic parts are subjected to a time-varying field. This can happen in

– brackets and welding plates/tension bars
– collars
– end plates and mirror plates
– pins, keys and rods
– shells (helium containment for example) and shields etc.

Of course, closed flux loops as would be formed by welding seams on the pole surface are strictly forbidden.

Figure 41 shows the cold mass of an LHC dipole including its mechanical structure.

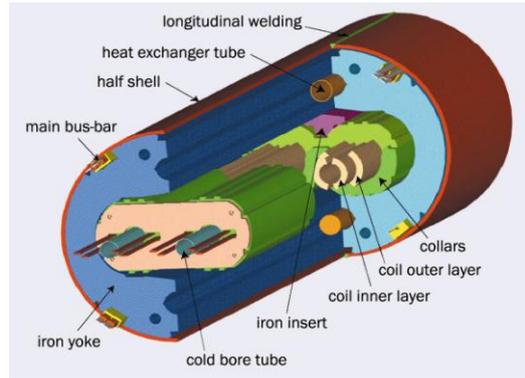

**Fig. 41:** Cold mass of the LHC dipole with structural parts

Very often eddy currents occur only at high fields, when, because of saturation, flux leaks out of the iron yoke. For example, for the R&D magnet GSI001 the nominal field was increased from 3.5 T to 4 T.

Figure 42 (a) demonstrates that at this high saturation level the field leaks and induces eddy currents in the copper shield. Figure 42 (b) gives the power density for a ramp rate of 4 T/s. The maximum value is 270 kW/m$^3$ [24].

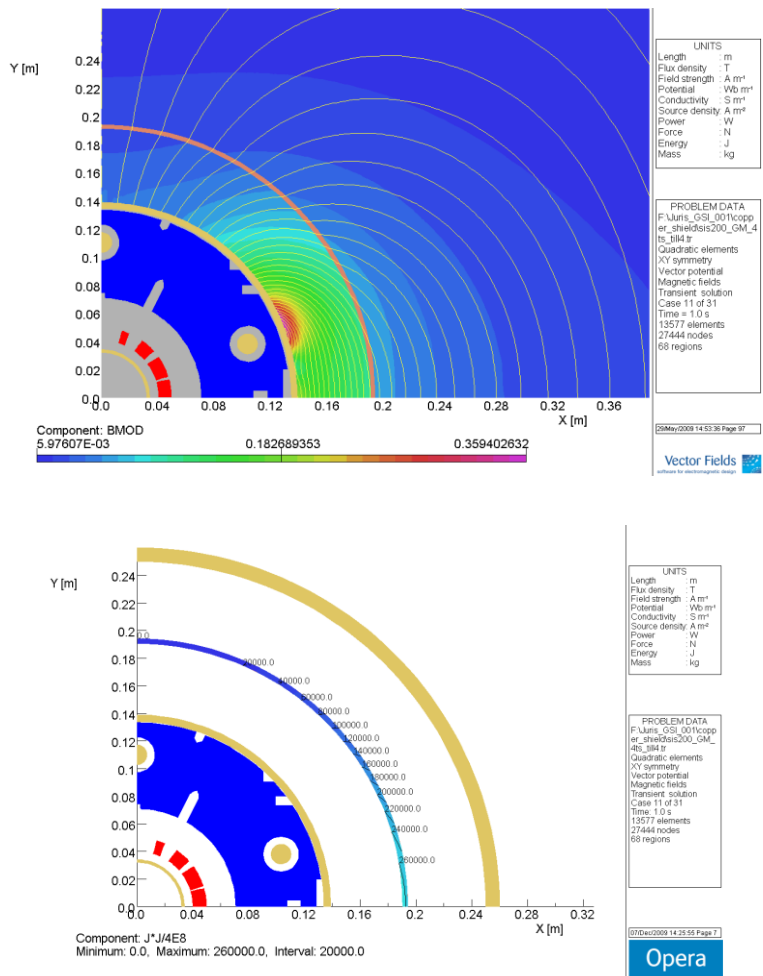

**Fig. 42**: (a) $B_{mod}$ at 4T nominal bore field and (b) corresponding power loss density in the copper shield at 4 T/s

Collars and iron laminations are held together by rods, pins and keys. They usually form a flux loop. Rods should be insulated from the iron by insulating sleeves and washers. If that is not possible for mechanical reasons, they have to be placed in positions where the corresponding flux is minimal (Fig. 43) [21].

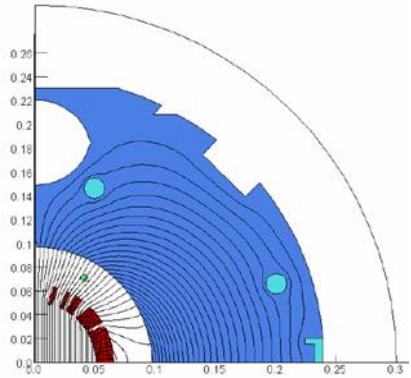

**Fig. 43:** Optimal position of the rods (flux enclosed between symmetric rods is minimized)

Figure 44 shows the eddy current density in rods and keys of the SIS 300 dipole prototype, calculated with ELEKTRA$^{©}$. The iron key shows a contribution only at the end, the central part 'sees' no flux change [21].

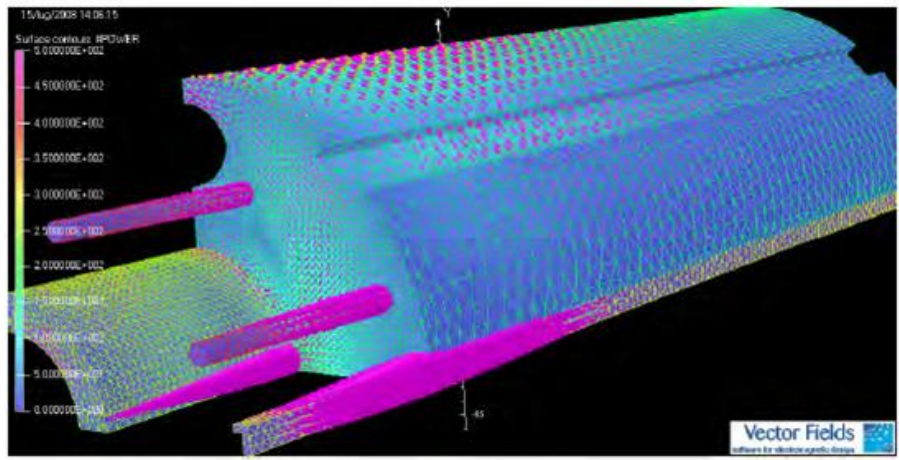

**Fig. 44:** Eddy current density in rods and keys of the mechanical structure of the SIS 300 dipole prototype

### 3.4 Eddy currents in beam pipe

The analytical formulas for a round beam pipe were given in Eq. (37).

Here we discuss elliptical beam pipes for the special case of a pipe at 4 K, as part of the cold mass of the synchrotron dipole of the planned synchrotron SIS 100. In Fig. 45 is shown the eddy current loss density in watts per metre of an infinitely long stainless-steel pipe at 4 K as the result of an ANSYS-calculation [25], [26]. The main loss occurs on both sides of the pipe. The wall thickness $d$ is 0.3 mm, the loss (averaged over the whole pipe) is 4.9 W/m for a ramp rate of 4 T/s. Here $a = 64$ mm and $b = 29$ mm are major and minor semi-axes of the elliptical cross-section.

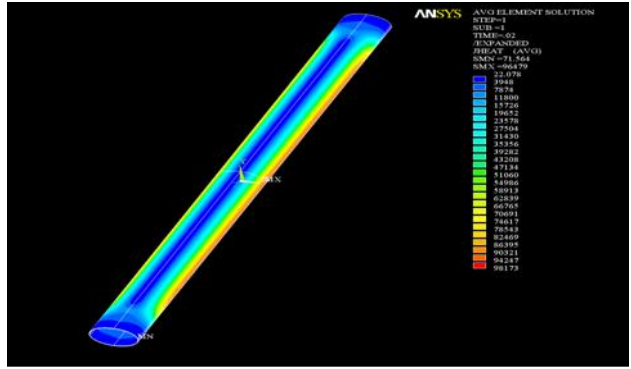

**Fig. 45:** Eddy current loss density distribution in an elliptical beam pipe at 4 K

An analytical formula can be derived for this simple 2D geometry (*l* length of the pipe) [27]:

$$P = \frac{a^2}{\rho} \dot{B}^2 f(\varepsilon) a\, d\, l \qquad (43)$$

with

$$\varepsilon = \sqrt{1 - (b/a)^2}$$

$$f(\varepsilon) = \int_0^{2\pi} \cos^2 \eta * \sqrt{1 - \varepsilon^2 \cos^2 \eta}\, d\eta\ .$$

Please note again the quadratic dependence on the 'relevant' dimension *a*, whereas the product $f(\varepsilon)adl$ is proportional to the volume.

The main reason for the use of a cold beam pipe is cryogenic pumping, especially of hydrogen. That requires that large parts of the pipe stay below 10 K. Of course, the heat produced by the eddy currents has to be carried away. One option is to use cooling pipes connected to the beam pipe with liquid helium flowing inside. It is clear that the design must avoid any flux loop coupled to the main field component (Fig. 46) [28].

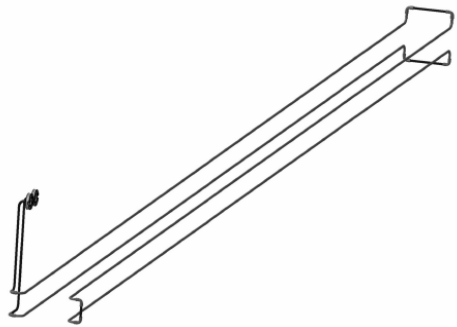

**Fig. 46:** Design of the cooling pipes of an elliptical beam pipe

In addition, reinforcement ribs have to be added to the tube (in order to guarantee the mechanical stability), leading to a real 3D problem. ANSYS calculations were made for the central and the end part of the beam pipe [29]. Figure 47 shows the results.

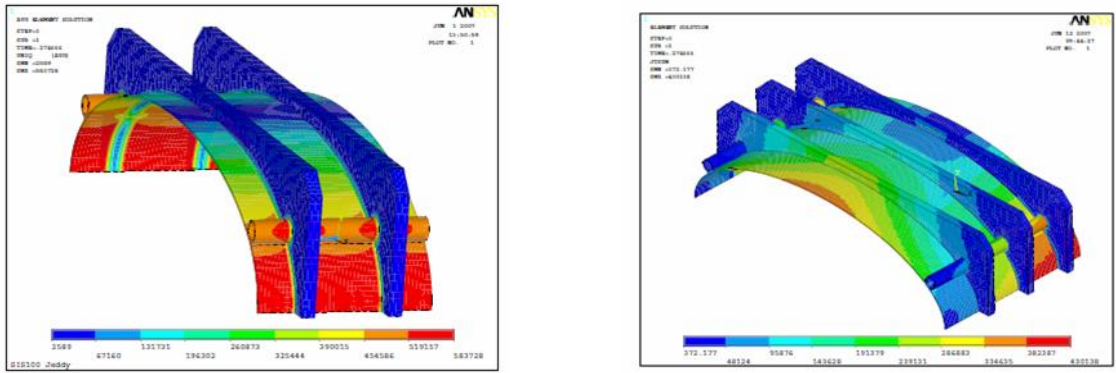

**Fig. 47:** Eddy current density distribution in the central (left) and end part of the beam pipe (right)

Unfortunately, the average loss in watts per metre is now increased from 4.9 W/m to 8.7 W/m. Electrical insulation of the cooling pipes helps to reduce the losses.

# 4    Magnet design principles

Based on the previous section we are now able to summarize the design principles for pulsed or ramped magnets, in order to minimize eddy current effects:

–   The magnet must be laminated. The lamination sheets must be well insulated. The thickness of the sheet is determined by the ramp rate or the frequency of the operation.

–   Iron with an increased resistance (mostly by increasing the silicon content) should be used. However, more important is a low coercivity.

–   The magnet design must be appropriate:

    –   Saturation should be avoided in the magnet-end field region. Owing to the anisotropy and non-linearity of the permeability, the longitudinal field component penetrates further into the iron at high fields (static phenomenon!), creating more eddy currents in this end region in case of a time-varying field (see 3.1.2.1.3).

    –   Saturation should also be avoided in the central part of the magnet, because flux leakage may induce eddy currents in structural components [30].

    –   A Rogowski or a simple chamfer pole profile at the magnet pole ends reduces the power loss, compared to a 'sharp edge', since the longitudinal magnetic field component perpendicular to the lamination sheets comes down.

    –   Vertical or horizontal slits in the end laminations limit the 'relevant dimension' of the eddy currents and therefore reduce the effects [31]. Be aware of problems with mechanical stability and field quality.

    –   Non-conductive pole material at the magnet ends prevents eddy currents, but reduces field quality.

    –   'Long' magnets are advantageous, also from the eddy current perspective, since most of the eddy current contributions come from the magnet ends.

    –   Be aware that the coil shape at the magnet end has an influence on the horizontal field component.

    –   For superconducting coils special low-loss wire and cable must be used.

- The mechanical structure of the magnet must be carefully designed to avoid flux loops and flux concentration in bulky components, especially if they are made out of low-carbon iron. Stainless steel may be a better choice.
- Replace conductive materials by non-conductive materials wherever possible.

Field control is a very efficient way to reduce the waiting time for the eddy currents to die out [32].

## 5   Summary

This paper covers the main eddy current effects in accelerator magnets: field modification (time delay and field quality) and resistive power losses. In the first part, starting from the Maxwell equations, a basic understanding of the processes was given and explained with examples of simple geometry and time behaviour. The 'magnetic diffusion' approach was used and the 'small perturbation' method was explained. Useful formulas were derived for an analytic estimate of the size of the effects.

In the second part, the effects in real magnets were analysed and described with a comparison between numerical and measured results. It was demonstrated that eddy currents play a role in the yoke, coil, mechanical structure, and beam pipe of a magnet.

Finally, based on the previous main parts, design recommendations were given, regarding how to minimize eddy current effects.

**Acknowledgements**

The author wishes to thank Dr. S.Y. Shim for his help during the preparation of this paper and Prof. A. Kalimov for many helpful discussions.